\documentclass[final,3p,times,sort&compress]{elsarticle}

\usepackage{amssymb}
\usepackage{mhchem}
\usepackage{setspace}
\usepackage{lineno}
\usepackage{enumitem}
\usepackage{subfigure}
\usepackage{float}
\usepackage[T1]{fontenc}
\usepackage{graphicx,multirow,booktabs,caption}
\usepackage{booktabs,multirow,array}
\usepackage[colorlinks,citecolor=blue,filecolor=blue,linkcolor=blue,urlcolor=black,breaklinks=ture,linktocpage=ture,plainpages=false]{hyperref}
\begin{document}
\doublespacing

\begin{frontmatter}

\title{Predicting grain boundary energies of complex alloys from ab initio calculations}

\author[label1]{Changle Li}

\author[label1]{Song Lu\corref{cor1}}
\ead{songlu@kth.se}

\author[label1,label3,label4]{Levente Vitos}

\address[label1]{Applied Materials Physics, Department of Materials Science and Engineering, KTH Royal Institute of Technology, SE-10044 Stockholm, Sweden}

\address[label3]{Department of Physics and Astronomy, Division of Materials Theory, Uppsala University, Box 516, SE-75120 Uppsala, Sweden}

\address[label4]{Research Institute for Solid State Physics and Optics, Wigner Research Center for Physics, P.O. Box 49, H-1525 Budapest, Hungary}

\cortext[cor1]{Corresponding author}

\begin{abstract}
Investigating the grain boundary energies of pure fcc metals and their surface energies obtained from~\textit{ab initio} modeling, we introduce a robust method to estimate the grain boundary energies for complex multicomponent alloys. The input parameter is the surface energy of the alloy, which can easily be accessed by modern~\textit{ab initio} calculations based on density functional theory. The method is demonstrated in the case of paramagnetic Fe-Cr-Ni alloys for which reliable grain boundary data is available. 
\end{abstract}



\end{frontmatter}


Grain boundary (GB) plays a critical role in the microstructural evolution of polycrystalline materials~\cite{sutton1995interfaces, rohrer2010comparing}. 
The structures and energetics of GBs are closely related to various physical and mechanical properties, e.g., alloying segregation, precipitation, coarsening, or crack. 
Particularly, tailoring the properties of GBs by, e.g., controlling the element segregation/depletion, has been an important strategy for improving mechanical properties of alloys, and it is often referred to as the 'grain boundary engineering'~\cite{randle2004twinning, randle2006mechanisms}.
The GB geometry is described by the five degrees of freedom (DOFs).
For a complete description, three degrees are assigned to the vector relating the misorientation of two adjoining grains and the other two describe the GB inclination plane~\cite{krause2019review}. 
As a key feature, the GB energy (GBE, or $\gamma_{\rm GB}$) varies significantly with both misorientation and inclination~\cite{rohrer2011grain, bulatov2014grain}, which makes accurate determination of the GBEs a great challenge in both experimental measurements and computational simulations.

Experimentally, the mean GBE can be estimated from the geometries of the surface triple junctions or internal GB triple junctions in zero-creep experiments~\cite{herring1951physics}. 
Often, it is the relative GBE or the GBE anisotropy that are determined~\cite{, kudrman1969relative, smith1948grains}. 
For example, measuring the groove angle at the equilibrium junction composed of grain boundary and surface, the GBE to surface energy ratio can be determined~\cite{gjostein1959absolute, hasson1971interfacial, mykura1955interferometric}. 
Assuming that the surface energy is known, the GBE can be determined.
Similar procedure has been applied for the evaluation of the low- and high-angle GBEs, as well as the twin boundary energies~\cite{murr1970interfacial, roth1975surface, bolling1968average, murr1973twin}. 
Unfortunately, the surface energy itself is a quantity difficut to determine accurately by experiments, especially at low temperatures. 
Most of the available room temperature experimental surface energy were obtained indirectly by extrapolating the surface tension measured in the liquid phase~\cite{tyson1977surface, de1988cohesion, vitos1998surface}. 
The GBE can also be determined using diffusion data. 
Borisov~\cite{borisov1964relation} proposed a semi-empirical relationship between the increase of the self-diffusion in the GB relative to the bulk and the absolute GBE. 
This method was examined by Pelleg~\cite{pelleg1966relation} and a satisfactory agreement between directly measured GBEs and the calculated ones was reported for a few close-packed metals.
The same method has been applied to alloys (e.g., Au-Ta alloys~\cite{gupta1976grain, gupta1977influence} and Ni alloys~\cite{prokoshkina2013grain}).
Despite the existing experimental methods for GBE determination, it is cumbersome to perform these measurements considering the large number of GB types and their composition, temperature, and magnetic state dependences.

Alternatively, GBs can be investigated by atomistic simulations using empirical potentials or density functional theory (DFT) calculations. 
Potentials using embedded atomic method (EAM) applied for the GB studies in pure metals~\cite{rittner1996110, wolf1989structure, wolf1989role, udler1996grain, shiga2003structure, tschopp2007asymmetric} led to improved understanding of the atomic structures and their energetics. 
For example, Holm~\textit{et al.}~\cite{holm2010comparing, olmsted2009survey} calculated the GBEs for a large number of GBs in pure face-centered cubic (fcc) metals and found that the GBEs in different materials are strongly correlated. 
Olmsted~\textit{et al.}~\cite{olmsted2009survey} showed that the GBEs are more influenced by the grain boundary plane than the misorientation angle. 
Studies based on DFT are usually more accurate and have a better predictive power as compared to methods based on empirical potentials. On the other hand, DFT calculations are often limited to pure metals or simple alloys and to low-index coincidence site lattice types of GBs due to the extensive computational burden~\cite{zheng2020grain, scheiber2016ab, wang2018grain}.

In the present work, we adopt DFT calculations to study the correlation between the GBEs of various fcc metals and between the GBEs and the surface energy. 
We show that the GBEs in a pair of fcc metals are strongly correlated via a material dependent factor ($\delta$). This parameter can be estimated from the ratio of the low-index surface energies. 
Considering that DFT methods for studying the surfaces of pure metals~\cite{lee2018surface, swart2007surface} and alloys~\cite{pitkanen2013ab} have readily been established, the present development puts forward a robust method for predicting the GBEs of complex alloys using their surface energies.

Ten types of the $[1\bar{1}0]$ tilt GBs in ten fcc metals (X=Al, Cu, Au, Ag, Ni, Pd, Pt, Co, Rh, and Ir) were calculated by the Vienna~\textit{ab initio} Simulation Package (VASP)~\cite{kohn1965self} using the projector augmented wave (PAW) method~\cite{blochl1994projector}. 
The atomic structures of these GBs are presented in Table S1 and Fig. S1 in the Supplementary Material (SM). 
For the exchange-correlation funtional we adopted the generalized gradient approximation parameterized by the Perdew, Burke, and Ernzerholf (PBE)~\cite{perdew1996generalized}. 
The~\textit{k}-point meshes were carefully tested to ensure the convergence of the GBEs within $\sim$0.02 J/m$^{2}$. 
Cutoff energies were set to 500 eV for all metals. 
Full geometry relaxation was performed and the convergence criteria for electronic energy and force calculations were $10^{-5}$ eV and 0.02 eV/\AA, respectively.

\begin{figure}[ht!]
	\centering
	\subfigure[]{\label{fig:01a}\includegraphics[scale=0.4]{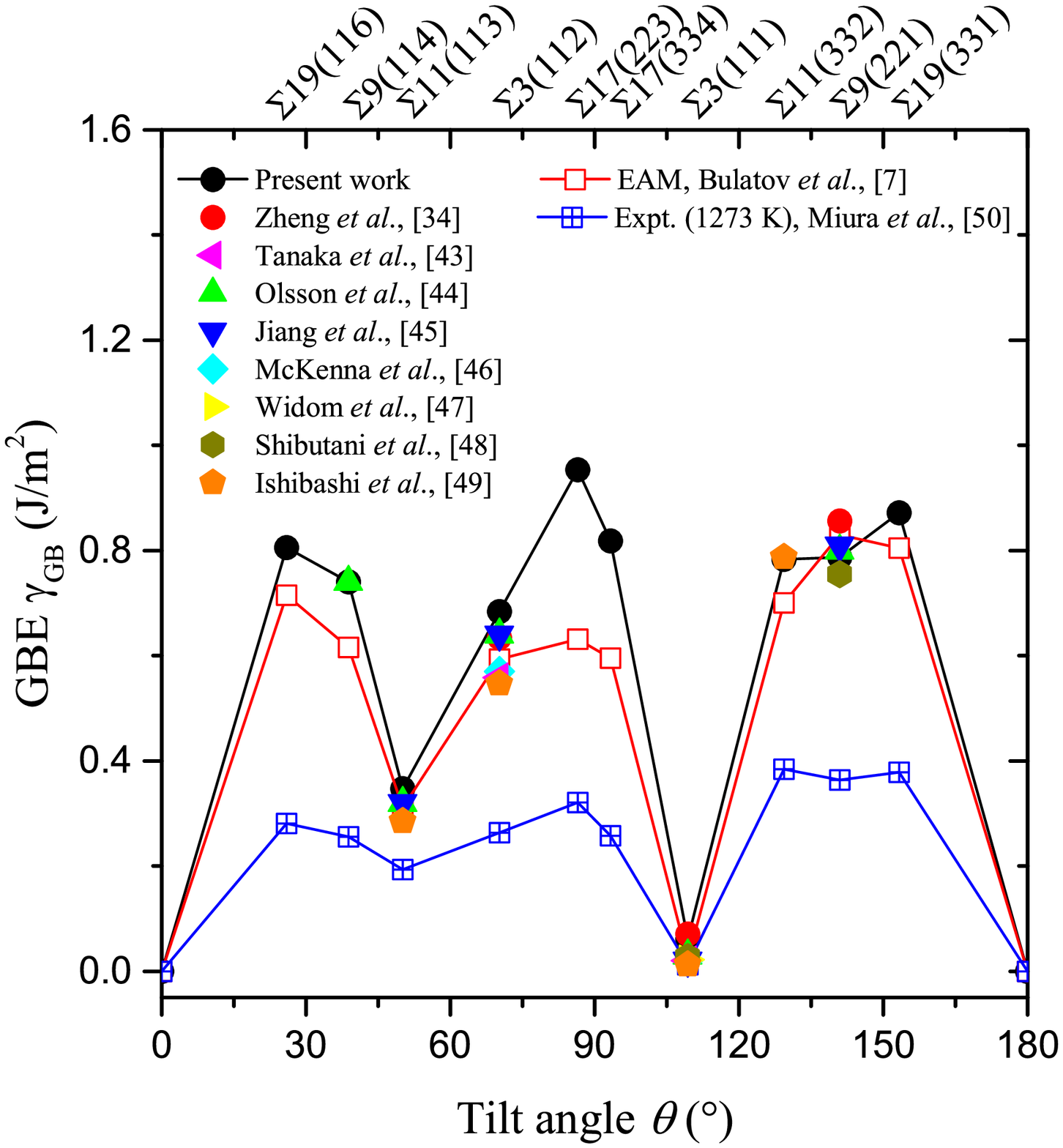}}
	\hspace{0.1in}
	\subfigure[]{\label{fig:01b}\includegraphics[scale=0.4]{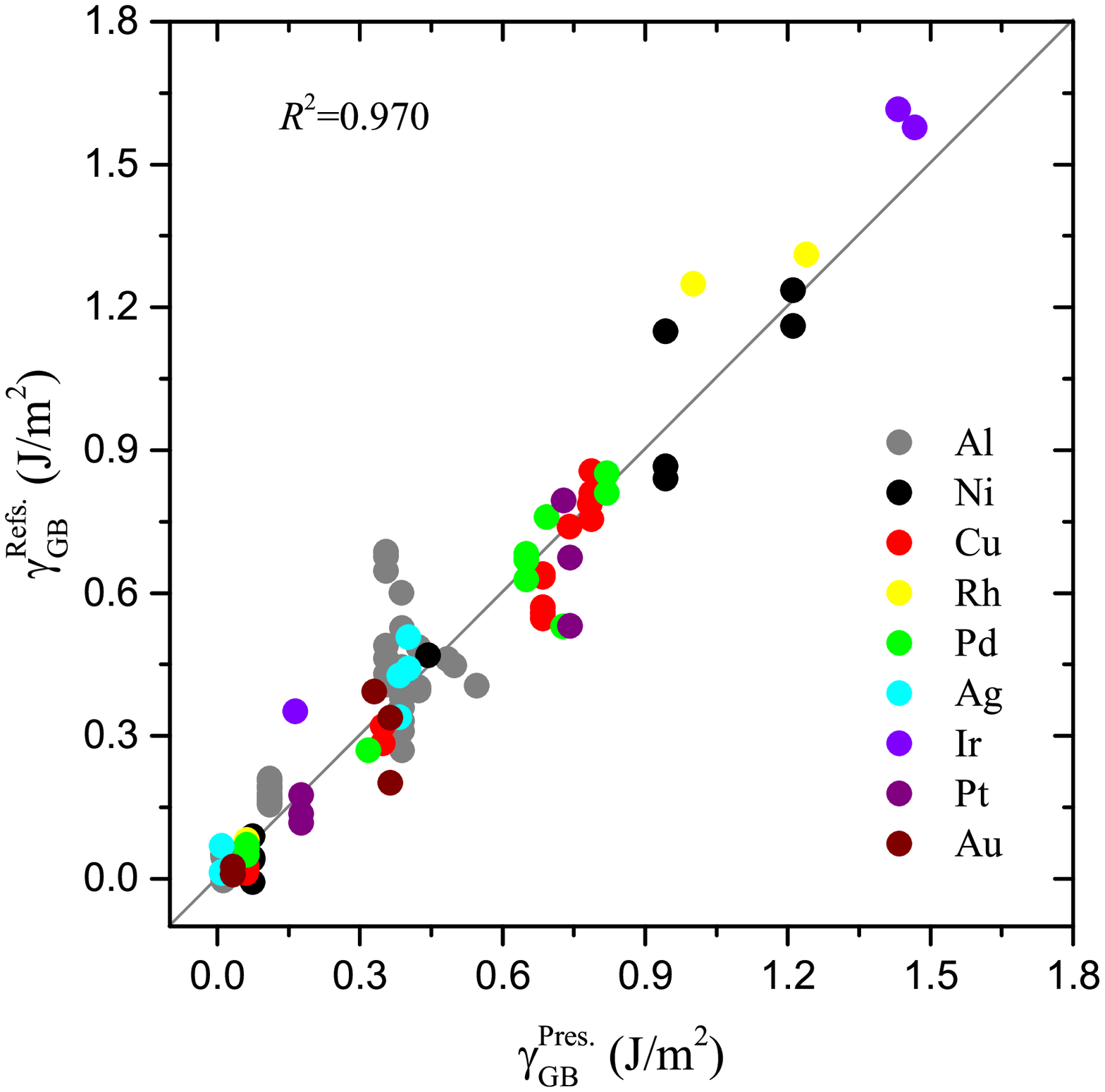}}
	\caption{\label{fig:01}(Color online) Theoretical and experimental GBEs for ten selected fcc metals. \label{fig01-a}(a) Comparison between the present and previous $\gamma_{\rm GB}$ values for Cu as a function of the $[1\bar{1}0]$ tilt angle $\theta$. The previous DFT (solid symbols)~\cite{zheng2020grain, nishiyama2020application, hallberg2016investigation, xu2016self, bean2016origin, gao2014first, tsuru2010incipient, wang2009first}, EAM (squares)~\cite{bulatov2014grain} and the high temperature experimental (squares with plus)~\cite{miura1994temperature} data are indicated. On the top of the figure, we show the ten GBs for which the present~\textit{ab initio} calculations were performed. (b) Comparison between the present ($\gamma_{\rm GB}^{\rm Pres.}$) and previous ($\gamma_{\rm GB}^{\rm Refs.}$)~\cite{zheng2020grain, hallberg2016investigation, xu2016self, bean2016origin, gao2014first, tsuru2010incipient, wang2009first, uesugi2011first, mahjoub2018general, nishiyama2020application, yamaguchi2019first, inoue2007first, tsuru2009fundamental, cao2018correlation, pang2012mechanical, janisch2010ab, wright1994density, thomson1997ab, thomson2000insight, chen2017role, pan2018development, siegel2005computational, lovvik2018grain, o2018grain} DFT results obtained for the ten selected metals (shown in the legend) and ten tilt angles (not shown). Numerical values are listed in Table S2 in SM. The $R^{2}$ value obtained for the linear fit considering all selected metals shows a high degree of correlation.}
\end{figure}

In Fig.~\ref{fig:01a}, we show the calculated $\gamma_{\rm GB}$ for Cu as a function of the $[1\bar{1}0]$ tilt angle $\theta$, in comparison with the previous DFT values~\cite{zheng2020grain, nishiyama2020application, hallberg2016investigation, xu2016self, bean2016origin, gao2014first, tsuru2010incipient, wang2009first} as well as the EAM results~\cite{bulatov2014grain}. 
The available experimental GBEs at high temperature are also included. 
The DFT results agree well with each other, showing the typical shape of the $\gamma$-surface for the $[1\bar{1}0]$ tilt GBs with two energy minima located at $\Sigma$3(111) (corresponding to $\theta=109.47$\textdegree) and $\Sigma$11(113) ($\theta=50.48$\textdegree).
The EAM and experimental $\gamma$-surfaces show the same shape~\cite{bulatov2014grain, miura1994temperature} although the absolute values are somewhat smaller than the DFT results. 
We associate the deviations between the DFT and experimental values to thermal effects, which were neglected in the DFT calculations, and to the fact that the experimental values are relative results~\cite{miura1994temperature}. 
In Fig.~\ref{fig:01b}, we compare the present GBEs for all fcc metals with the available DFT results~\cite{zheng2020grain, hallberg2016investigation, xu2016self, bean2016origin, gao2014first, tsuru2010incipient, wang2009first, uesugi2011first, mahjoub2018general, nishiyama2020application, yamaguchi2019first, inoue2007first, tsuru2009fundamental, cao2018correlation, pang2012mechanical, janisch2010ab, wright1994density, thomson1997ab, thomson2000insight, chen2017role, pan2018development, siegel2005computational, lovvik2018grain, o2018grain} (numerical values are listed in Table S2 in SM). 
Overall, the present GBEs have an excellent agreement with the former DFT values, with $R^{2}=0.97$ and a standard error of $\sim$0.017 J/m$^{2}$.

\begin{figure}[ht!]
	\centering
	\includegraphics[scale=0.5]{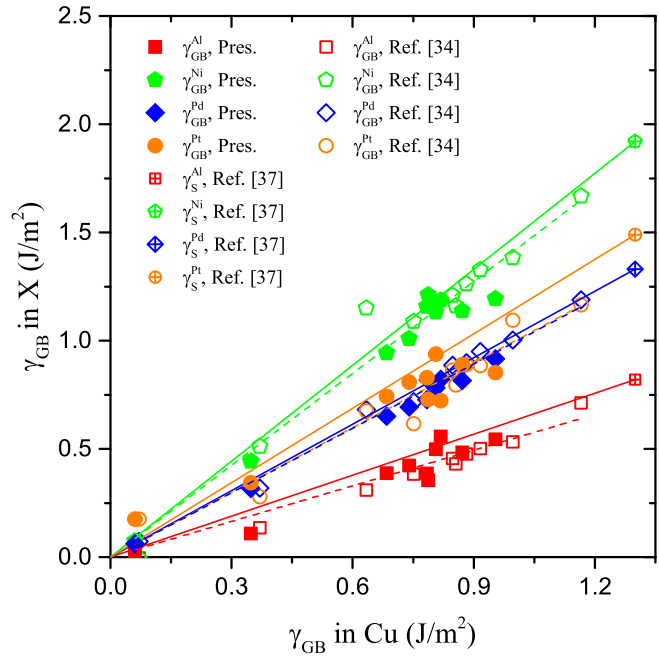}
	\caption{\label{fig:02}(Color online) Pairwise comparison of the calculated $\gamma_{\rm GB}$ (solid symbols) and the (111) surface energies (open symbols with plus) for Al, Ni, Pd, and Pt with Cu obtained at 0 K. The dashed lines are the linear fits to the GBEs. Previous DFT GBEs (open symbols) for tilt and twist GBs from Ref.~\cite{zheng2020grain} are included in the linear fitting. The DFT surface energies are taken from Ref.~\cite{lee2018surface}.}
\end{figure}

Starting from the present DFT results for the GBEs, we examine the correlation between the GBEs in different metals. 
Here, we choose Cu as the reference and compare the GBEs of the same GB structure in different materials. 
We notice however that choosing another metals as reference leads practically to the same conclusions. 
Results for four elements, Al, Ni, Pd, and Pt, taken as examples of $sp$, $3d$, $4d$, and $5d$ metals, respectively, are shown in Fig.~\ref{fig:02}. 
In order to strengthen our point, some previous DFT results for tilt and twist GBs~\cite{zheng2020grain} are also included in the figure. 
We emphasize that the following observations and discussions apply to all metals considered here and the detailed results can be found in SM (Fig. S2 and Table S2). 
First, we confirm that there is a clear correlation between the GBEs in different metals, as demonstrated by the previous EAM results~\cite{ratanaphan2015grain}. 
All the boundary energies in a specific metal locate approximately on a straight line passing through the origin (dashed line in Fig.~\ref{fig:02}), indicating that a single material dependent factor ($\delta$) may be used to correlate the GBEs in a pair of metals, i.e., $\gamma_{\rm GB}^{\rm A}(\rm DOF)\approx\delta \gamma_{\rm GB}^{\rm Cu}(\rm DOF)$ (A stands for an fcc metal). 
The slopes ($\delta_{\rm GB(fit)}^{\rm A/Cu}$) of the linear fitting for all 9 metals are listed in Table~\ref{table1}. 
The nearly perfect scaling relation between the GBEs in different metals highlights the critical roles played by the GB structures in deciding the GBEs. 
The anisotropy of GBEs is, to the first order approximation, decided by the boundary structure, i.e., the five DOFs. 
In other words, despite of the existence of the difference in local atomic structure configuration (or even in magnetic environment), the GBEs in different materials may be described by an universal functional of the five geometric parameters in different materials. 
This observation follows closely the concept developed by Bulatov~\textit{et al.}~\cite{bulatov2014grain}.

\begin{table}[ht!]
	\centering
	\caption{\label{table1}Surface energies for the (100) and (111) surface facets ($\gamma_{\rm S(100)}$ and $\gamma_{\rm S(111)}$, respectively) and their ratios relative to that of Cu for the selected fcc metals. $\delta_{\rm GB(fit)}^{\rm A/Cu}$ is the gradient of the linear fitting of the DFT GBEs in Fig.~\ref{fig:02} and Fig. S2. $\gamma_{\rm GB}$ ($\delta_{\rm S (100)}^{\rm A/Cu}$) and $\gamma_{\rm GB}$ ($\delta_{\rm S (111)}^{\rm A/Cu}$) are the predicted GBEs using the (100) and (111) surface energies, respectively. $\gamma_{\rm GB}^{\rm expt.}$ are the experimental GBEs (references indicated). All the DFT calculations correspond to the static state (0 K), whereas the experimental GBEs for the general GBs are obtained by linear extrapolation to 0 K (see Fig. S3 in SM).}
	\begin{tabular}{l l l l l l | l l | l}
		\hline
		& \multicolumn{5}{c}{DFT} & \multicolumn{2}{c}{Predicted} & \multicolumn{1}{c}{Expt.} \\
		\hline
		& $\gamma_{\rm S(100)}$ & $\gamma_{\rm S(111)}$ & \multirow{2}*{$\delta_{\rm S (100)}^{\rm A/Cu}$} & \multirow{2}*{$\delta_{\rm S (111)}^{\rm A/Cu}$} & \multirow{2}*{$\delta_{\rm GB(fit)}^{\rm A/Cu}$ ($R^{2}$)} & $\gamma_{\rm GB}$ ($\delta_{\rm S (100)}^{\rm A/Cu}$) & $\gamma_{\rm GB}$ ($\delta_{\rm S (111)}^{\rm A/Cu}$) & $\gamma_{\rm GB}^{\rm expt.}$ \\
		& $(\rm J/m^{2})$ & $(\rm J/m^{2})$ & & & & $(\rm J/m^{2})$ & $(\rm J/m^{2})$ & $(\rm J/m^{2})$ \\
		\hline
		Cu & 1.44$^{a}$ & 1.30$^{a}$ & - & - & - & - & - & 0.78$^{c}$ \\
		Al & 0.92$^{a}$ & 0.82$^{a}$ & 0.64 & 0.63 & 0.56 (0.980) & 0.50 & 0.49 & 0.44$^{d}$ \\
		Au & 0.86$^{a}$ & 0.71$^{a}$ & 0.60 & 0.55 & 0.49 (0.989) & 0.46 & 0.42 & 0.40$^{e}$ \\
		Ag & 0.84$^{a}$ & 0.76$^{a}$ & 0.59 & 0.58 & 0.57 (0.997) & 0.45 & 0.45 & 0.45$^{f}$ \\
		Ni & 2.22$^{a}$ & 1.92$^{a}$ & 1.54 & 1.48 & 1.38 (0.995) & 1.20 & 1.15 & 1.11$^{g}$ \\
		Pd & 1.51$^{a}$ & 1.33$^{a}$ & 1.05 & 1.02 & 0.96 (0.998) & 0.81 & 0.79 & 0.80$^{h}$ \\
		Pt & 1.85$^{a}$ & 1.49$^{a}$ & 1.28 & 1.15 & 1.01 (0.987) & 1.00 & 0.89 & 0.84$^{i}$ \\
		Rh & 2.35$^{a}$ & 2.01$^{a}$ & 1.63 & 1.55 & 1.71 (0.994) & 1.27 & 1.20 & - \\
		Ir & 2.84$^{a}$ & 2.06$^{a}$ & 1.97 & 1.58 & 2.12 (0.994) & 1.53 & 1.23 & - \\
		Co & 2.46$^{b}$ & 2.02$^{b}$ & 1.71 & 1.55 & 1.45 (0.990) & 1.53 & 1.21 & - \\
		\hline
	\end{tabular} \\
	$^{a}$ Ref.~\cite{lee2018surface},\\
	$^{b}$ Ref.~\cite{swart2007surface},\\
	$^{c}$ Ref.~\cite{gupta1977influence},\\
	$^{d}$ Refs.~\cite{murr1973twin, bolling1968average, prokofjev2019estimation},\\
	$^{e}$ Ref.~\cite{gupta1977influence},\\
	$^{f}$ Refs.~\cite{bolling1968average, prokofjev2019estimation},\\
	$^{g}$ Refs.~\cite{divinski2010grain, murr1970interfacial, miedema1978surface}.\\
	$^{h}$ Refs.~\cite{miedema1978surface, birringer2002interface, birringer2003interface, prokofjev2019estimation}.\\
	$^{i}$ Refs.~\cite{mclean1971study, hondors1982metallic}.\\
\end{table}

\begin{figure}[H]
	\centering
	\includegraphics[scale=0.5]{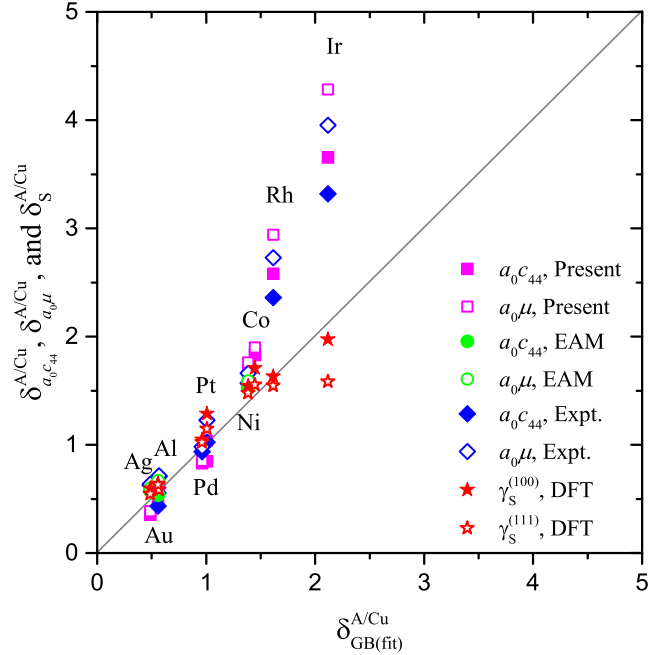}
	\caption{\label{fig:03}(Color online) Comparison of $\delta_{\rm GB(fit)}^{\rm A/B}$ with $\delta_{a_{0}c_{44}}^{\rm A/B}$, $\delta_{a_{0}\mu}^{\rm A/B}$, and $\delta_{\rm S}^{\rm A/B}$. The EAM and experimental shear moduli are from Refs.~\cite{holm2010comparing, simons1977single}. The surface energies are from Refs.~\cite{lee2018surface, swart2007surface}. Numerical values are listed in Table S5.}
\end{figure}

In literature, the mean GBEs for general GBs were proposed to scale with physical parameters like shear modulus ($a_{0}c_{44}$) or Voight average shear modulus ($a_{0}\mu$), cohesive energy ($E_{0}/a_{0}^{2}$), stacking fault energy (SFE, $\gamma_{\rm SF}$) or their combinations~\cite{holm2010comparing, ratanaphan2015grain, zheng2020grain}. 
The shear moduli and cohesive energy were rescaled by the lattice parameter ($a_{0}$) to give the same units as the GBE. 
The rationale behind the relation between the GBE and shear modulus~$\mu$, e.g., $\gamma_{\rm GB}\approx ka_{0}\mu$ ($k$, a coefficient) is from the Read-Shockley type of dislocation model~\cite{read1950dislocation}. 
There, the GBs with low misorientation angles are considered to be composed of arrays of dislocations whose energies are proportional to the shear modulus~\cite{holm2010comparing}. 
Therefore, the material dependent scaling factor $\delta$ that connects the GBEs in two materials is thought to be related to the ratio of $a_{0}c_{44}$ or $a_{0}\mu$ (denoted as $\delta_{a_{0}c_{44}}^{\rm A/B}$ and $\delta_{a_{0}\mu}^{\rm A/B}$ in the following). 
Holm~\textit{et al.}~\cite{holm2010comparing} showed that the ratios of both $a_{0}c_{44}$ and $a_{0}\mu$ are very close to the actual slope of the linear fit of the EAM GBEs for metals with low SFEs; while for metals with high SFEs like Al, the ratio of $a_{0}c_{44}$ acts as a better scaling factor than that of $a_{0}\mu$.
However, Ratanaphan~\textit{et al.}~\cite{ratanaphan2015grain} reported that the ratio of the cohesive energies ($\delta^{\rm A/B}_{E_{0}/a_{0}^{2}}$) is much better indicator than the ratios of $a_{0}c_{44}$ or $a_{0}\mu$ in bcc metals. 
It is argued that the GBEs scale with the cohesive energy based on the broken bond model of GBE~\cite{wolf1990broken}. 
But in fcc metals, EAM results~\cite{holm2010comparing, udler1996grain} indicate that the broken bond model does not give satisfactory prediction of GBEs, i.e., GBEs do not scale with $E_{0}/a_{0}^{2}$, nor one can use the ratio of $E_{0}/a_{0}^{2}$ to correlate the GBEs in different fcc metals.
DFT calculations also confirm that despite a general positive correlation between the GBE and the cohesive energy may exist, the overall correlation is weak~\cite{zheng2020grain}. 
As for the SFE, it is strongly related to the coherent twin boundary energy, $\gamma_{\rm SF}\approx2\gamma_{\rm tw}$, but correlates weakly with the general GBEs~\cite{holm2010comparing}. 
The above results indicate that the ratios of $a_{0}c_{44}$, $a_{0}\mu$, $E_{0}/a_{0}^{2}$ and SFE for a pair of materials are not likely to give a good prediction of $\delta$ that can be unambiguously used to correlate the GBEs in the randomly chosen fcc materials. 
In Fig.~\ref{fig:03}, we compare $\delta_{a_{0}c_{44}}^{\rm A/Cu}$ and $\delta_{a_{0}\mu}^{\rm A/Cu}$ with $\delta_{\rm GB(fit)}^{\rm A/Cu}$ in the studied metals with available DFT, EAM and experimental data (Table S3 in SM). 
Indeed, it shows that their agreement is strongly material dependent. 
For example, in Rh and Ir, $\delta_{a_{0}c_{44}}^{\rm A/Cu}$ and $\delta_{a_{0}\mu}^{\rm A/Cu}$ are about two times larger than $\delta_{\rm GB(fit)}^{\rm A/Cu}$. 

In high-index GBs, the geometry near the boundary is less close packed and many bulk-like bonds are missing which resembles locally a surface-like packing. 
Because of that, we consider the surface energy of close packed surface facets as an alternative indicator of the GBEs. 
Indeed, our analysis indicates that the ratio of the low-index surface energies ($\delta_{\rm S}^{\rm A/B}$) gives a highly accurate prediction of $\delta$. 
In Fig.~\ref{fig:02}, the (111) surface energies of pure metals locate approximatively on the same lines as the GBEs. 
For Al, Au, Ag, Ni, Pd, Pt, Co, and Rh, the ratios of the (111) surface energies ($\delta_{\rm S (111)}^{\rm A/Cu}$) are very close the actually slopes of the GBEs ($\delta_{\rm GB(fit)}^{\rm A/Cu}$) with mean deviation of $\sim$9$\%$. 
The largest overestimation is found for Ir, $\sim$25\%, which is still much better than the prediction based on shear moduli or cohensive energy (Fig.~\ref{fig:03}).  
Similar observations apply when using the (001) surface energies, see Table~\ref{table1}.

\begin{figure}[ht!]
	\centering
	\includegraphics[scale=0.5]{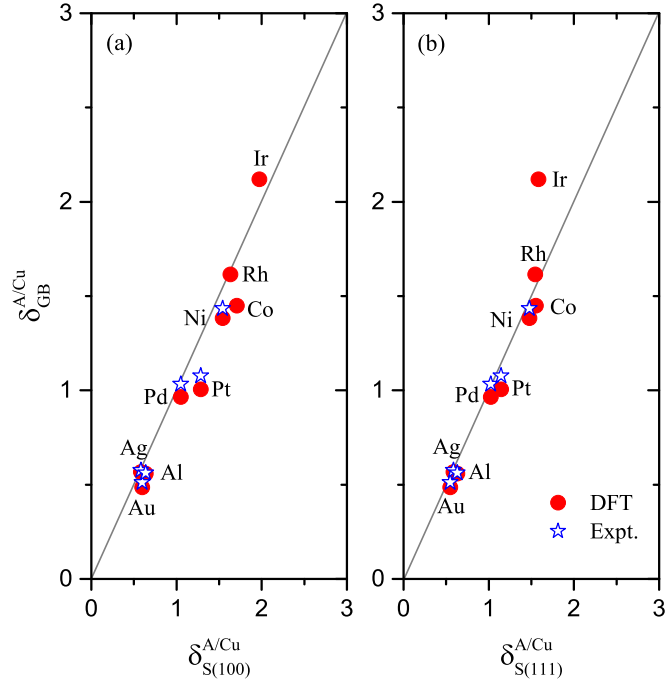}
	\caption{\label{fig:04}(Color online) Comparison of the ratio of the GBEs with the ratios of the low-index surface energies, (a) (100) surface and (b) (111) surface, for fcc metals. The ratios of the experimental GBEs (0 K) for the general GBs are also plotted.}
\end{figure}

We emphasize here that in the above analysis, both tilt and twist types of GBs are included. Therefore, we may anticipate that for the formation energies of the so called general GBs should also correlate with the surface energies.  
In Fig.~\ref{fig:04}, we compare the ratio of the experimental GBEs with the DFT calculated gradients ($\delta_{\rm GB(fit)}^{\rm A/B}$) and the ratio of the surface energies ($\delta_{\rm S (111)}^{\rm A/Cu}$ and $\delta_{\rm S (100)}^{\rm A/Cu}$, respectively). 
Indeed, a good agreement is reached for both (111) and (001) surface energies. 
The above result suggests an efficient approach for predicting the GBEs, especially the general GB for which the DOFs are not properly defined. 
We illustrate the approach by taking Cu as the reference system with measured general GBE of $\gamma_{\rm GB}^{\rm expt.Cu}$. 
Then the general GBE of an fcc metal A can be predicted as $\gamma_{\rm GB}^{\rm A}\approx\delta_{\rm S}^{\rm A/Cu}\times\gamma_{\rm GB}^{\rm expt. Cu}$ where $\delta_{\rm S}^{\rm A/Cu}$ is the ratio of the surface energies of metal A and Cu. 
This ratio changes very weakly with temperature up to the room temperature (see SM) and thus the formula is expected to apply at both low and room temperature. 
Following this approach, we predicted the 0 K GBEs of all metals considered here. 
Results are listed in Table~\ref{table1}. 
It turns out that the proposed surface energy-based scheme gives highly reliable predictions.

Most importantly, the same approach can be applied to complex alloys for predicting the general GBEs. 
Here we consider the paramagnetic stainless steel 304 ($\rm Fe_{0.71}Cr_{0.20}Ni_{0.09}$, atomic concentration) as an example, for which the (111) surface energy was calculated to be 2.83 $\rm J/m^{2}$ at 0 K~\cite{pitkanen2013ab}. 
Taking Cu as the reference system again, the ratio of the surface energy $\delta_{\rm S(111)}^{\rm Fe_{0.71}Cr_{0.20}Ni_{0.09}/Cu}$ is 2.18 and using the room-temperature general GBE of Cu (0.74 $\rm J/m^{2}$, Ref.~\cite{gupta1977influence}), for the general GBE of the 304 stainless steel we predict 1.61 $\rm J/m^{2}$. 
Our value agrees well with the experimental one of 1.67 $\rm J/m^{2}$  reported at room temperature by Murr~\textit{et al.}~\cite{murr1973measurement}, especially when considering the errors associated with the experimental values, such as the linear temperature dependence and the effects from minor alloying elements. 
In fact, with the concentration dependent surface energy calculated, we can also provide a parameterized function of predicting the GBE with respect to the chemical variations. 
Pitkänen \textit{et al.}~\cite{pitkanen2013ab} provided the regression function for the (111) surface energy with respect to the composition in ${\rm Fe}_{1-c-n}{\rm Cr}_{c}{\rm Ni}_{n}$ stainless steel, viz. 
\begin{equation}
	\gamma_{\rm S(111)}=1.279c-0.143n+2.588~(\rm J/m^{2}),
\end{equation}
where $c$ and $n$ are the atomic fractions of Cr and Ni contents, respectively. 
The above formula was established for compositions $0.12\leqslant c\leqslant 0.32$ and $ 0.04\leqslant n\leqslant 0.32$.
Now using the present approach based on surface energies, for the general GBE of ${\rm Fe}_{1-c-n}{\rm Cr}_{c}{\rm Ni}_{n}$ alloys we obtain 
\begin{equation}
	\gamma_{\rm GB}=0.727c-0.081n+1.471~(\rm J/m^{2}).
\end{equation}
The above linear expression in terms of weight percent (wt.\%) becomes 
\begin{equation}
	\gamma_{\rm GB}=0.00766x-0.00070y+1.474~(\rm J/m^{2}),
\end{equation}
where $x$ and $y$ are the weight percent of Cr and Ni contents, respectively. 
The variables are within the limits ($11\leqslant x\leqslant 30$, $4\leqslant y\leqslant 34$, wt.\%).

In summary, we explored the correlation between the GBEs in fcc metals with~\textit{ab initio} calculations. 
Our results demonstrated that the GBEs in fcc metals are strongly correlated, with a primary origin coming from the GB structure. 
A material dependent parameter $\delta$ is expected to scale the GBEs of the same GB structure in a pair of fcc metals. 
Here, we found that the ratio of the low-index surfaces can give a satisfactory estimation of $\delta$. 
Using~\textit{ab initio} surface energies and the reference data in Cu, we successfully predicted the general GBEs in other pure fcc metals and in a complex solid solution alloy. 
The present work introduces a feasible method for the prediction of the GBEs using~\textit{ab initio} calculations. 
We envision that with more~\textit{ab initio} studies for the GBs with structures varying in the space of the five DOFs in a reference metal, using the surface energy-based scaling parameters as proposed in the present work, the GBEs and anisotropy in complex alloys can be readily predicted. 

\section*{Acknowledgments}
The present work is performed under the project "SuperFraMat" financed by the Swedish Steel Producers' Association (Jernkontoret) and the Swedish Innovation Agency (Vinnova). 
The Swedish Research Council, the Swedish Foundation for Strategic Research, the Swedish Energy Agency, the Hungarian Scientific Research Fund, the China Scholarship Council, and the Carl Tryggers Foundation are also acknowledged for financial support. The computations were performed on resources provided by the Swedish National Infrastructure for Computing (SNIC) at the National Supercomputer Centre (NSC) in Linköping partially funded by the Swedish Research Council through grant agreement no. 2018-05973.





\bibliographystyle{elsarticle-num}
\bibliography{References.bib}

\end{document}


\begin{frontmatter}

\title{Supplementary Material for \\
	Predicting grain boundary energies of complex alloys from ab initio calculations}

\author[label1]{Changle Li}

\author[label1]{Song Lu\corref{cor1}}
\ead{songlu@kth.se}

\author[label1,label3,label4]{Levente Vitos}

\address[label1]{Applied Materials Physics, Department of Materials Science and Engineering, KTH Royal Institute of Technology, SE-10044 Stockholm, Sweden}

\address[label3]{Department of Physics and Astronomy, Division of Materials Theory, Uppsala University, Box 516, SE-75120 Uppsala, Sweden}

\address[label4]{Research Institute for Solid State Physics and Optics, Wigner Research Center for Physics, P.O. Box 49, H-1525 Budapest, Hungary}

\cortext[cor1]{Corresponding author}

\end{frontmatter}


\tableofcontents

\newpage
\section{GB structures}
\begin{figure}[H]
	\renewcommand{\thefigure}{S\arabic{figure}}
	\subfigure{\includegraphics[scale=0.5]{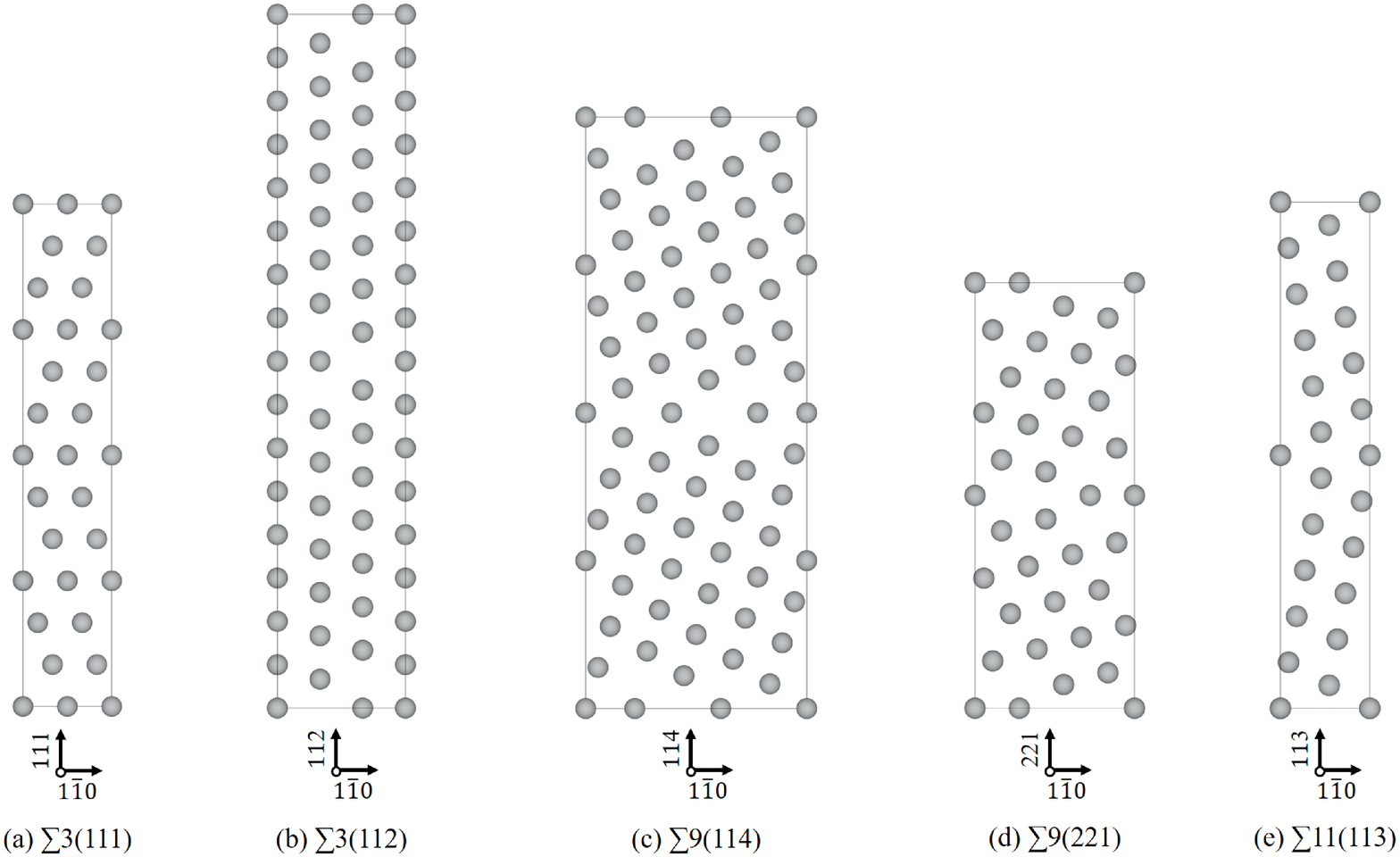}} \\
	\subfigure{\includegraphics[scale=0.5]{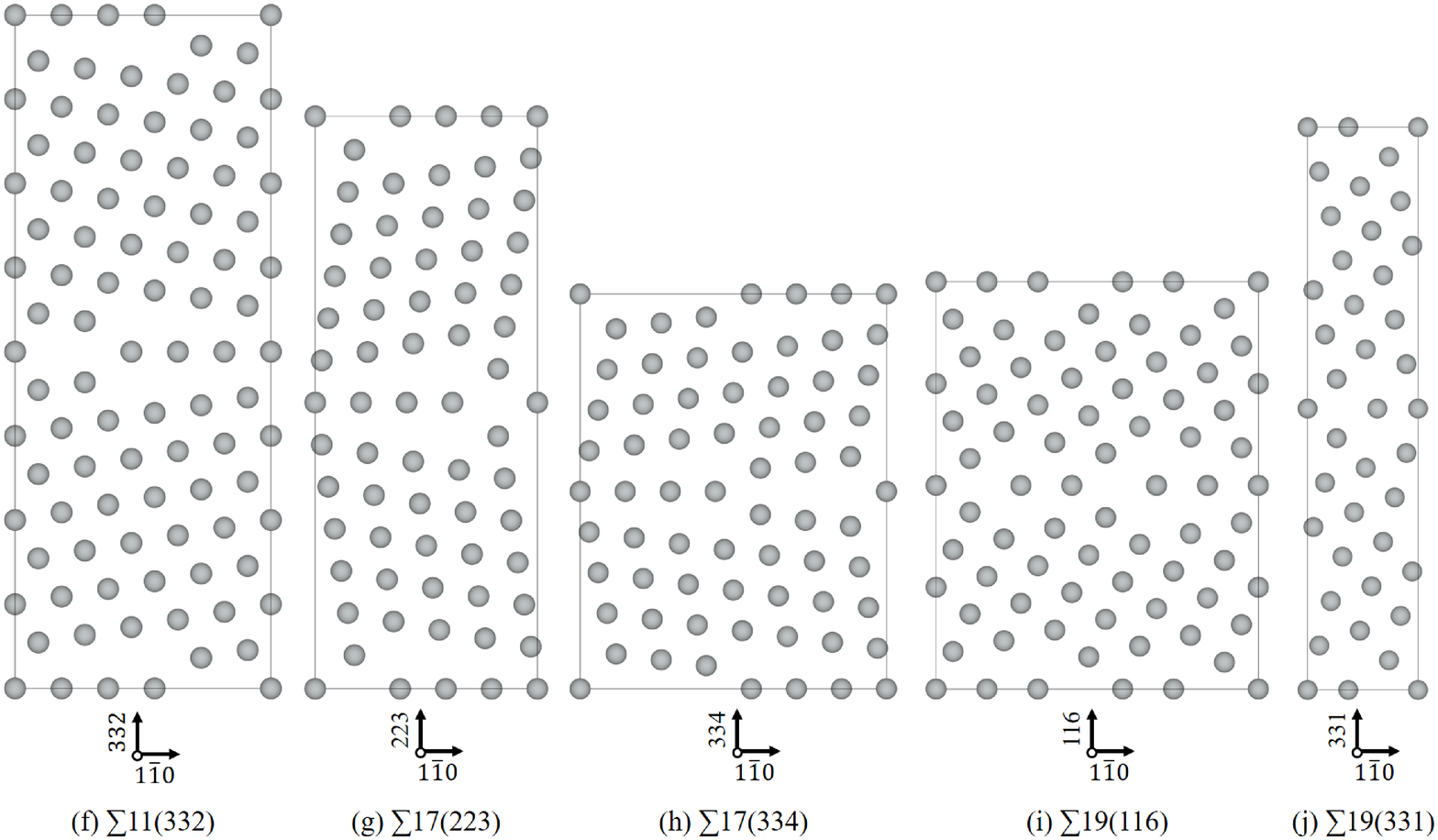}}
	\caption{\label{figS1}Schematics of the atomic GB structures before relaxation.}
\end{figure}

In the present work, we focus on symmetric tilt GBs with a $[1\bar{1}0]$ tilt axis. A set of 10 different GBs are listed in Table~\ref{tableS1}. All GBs are initiated from the tilt plane with certain misorientation angles. The schematics of the selected 10 GB structures are presented in Fig.~\ref{figS1}. 


\begin{table*}[h!]
	\centering
	\renewcommand{\thetable}{S\arabic{table}}
	\caption{\label{tableS1}The properties of the $[1\bar{1}0]$ tilt GBs studied in present study.}
	\begin{tabular}[1]{llll}
		\toprule
		Index & Angle & GB-plane & Number of atoms \\
		\midrule
		$\Sigma3$ & 109.47\ensuremath{^\circ} & (1 1 1) & 24 \\
		$\Sigma3$ & 70.53\ensuremath{^\circ} & (1 1 2) & 46 \\ 
		$\Sigma9$ & 38.94\ensuremath{^\circ} & (1 1 4) & 68 \\
		$\Sigma9$ & 141.06\ensuremath{^\circ} & (2 2 1) & 34 \\  
		$\Sigma11$ & 50.48\ensuremath{^\circ} & (1 1 3) & 22 \\ 
		$\Sigma11$ & 129.52\ensuremath{^\circ} & (3 3 2) & 82 \\  
		$\Sigma17$ & 86.63\ensuremath{^\circ} & (2 2 3) & 62 \\ 
		$\Sigma17$ & 93.37\ensuremath{^\circ} & (3 3 4) & 62 \\
		$\Sigma19$ & 26.53\ensuremath{^\circ} & (1 1 6) & 68 \\
		$\Sigma19$ & 153.35\ensuremath{^\circ} & (3 3 1) & 36 \\
		\bottomrule
	\end{tabular}
\end{table*}

\newpage
\section{The calculated GBEs}

The calculated GBEs for all studied metals are tabulated in Table~\ref{tableS2}. For comparison, the available DFT results in literature are also presented~\cite{zheng2020grain, uesugi2011first, mahjoub2018general, nishiyama2020application, yamaguchi2019first, inoue2007first, tsuru2009fundamental, cao2018correlation, pang2012mechanical, janisch2010ab, wright1994density, thomson1997ab, thomson2000insight, hallberg2016investigation, bean2016origin, gao2014first, tsuru2010incipient, wang2009first, chen2017role, pan2018development, siegel2005computational, lovvik2018grain, o2018grain}.

\begin{table}[H]
	{\small
		\renewcommand{\thetable}{S\arabic{table}}
		\caption{\label{tableS2}The calculated GBEs (in unit of $\rm J/m^{2}$) in comparison with available DFT results in literature.}
		\begin{tabular*}{1.17\textwidth}{llllllllllllll}
			\toprule 
			& $\Sigma$19(116) & $\Sigma$9(114) & $\Sigma$11(113) & $\Sigma$3(112) & $\Sigma$17(223) & $\Sigma$17(334) & $\Sigma$3(111) & $\Sigma$11(332) & $\Sigma$9(221) & $\Sigma$19(331) & Functional & Ref. \\ 
			\midrule 
			Al & 0.560 & 0.423 & 0.109 & 0.389 & 0.545 & 0.557 & 0.012 & 0.387 & 0.355 & 0.484 & PBE & Present \\
			& & & & 0.310 & & & -0.004 & & 0.430 & & PBE & \cite{zheng2020grain} \\
			& 0.448 & 0.402 & 0.179 & 0.379 & & & 0.045 & & & & PBE & \cite{uesugi2011first} \\
			& & & 0.161 & 0.527 & 0.406 & & 0.014 & & 0.646 & 0.462 & PBE & \cite{mahjoub2018general} \\
			& & & & 0.396 & & & 0.060 & & 0.463 & & PBE & \cite{nishiyama2020application} \\
			& & & 0.170 & 0.360 & & & 0.050 & & 0.490 & & PBE & \cite{yamaguchi2019first} \\
			& & & & & & & & & 0.408 & & PW91 & \cite{inoue2007first} \\
			& & & 0.166 & 0.332 & & & 0.051 & 0.445 & & & GGA & \cite{tsuru2009fundamental} \\
			& & & 0.156 & 0.385 & & & 0.051 & 0.600 & 0.687 & & LDA & \cite{cao2018correlation} \\
			& & 0.395 & 0.211 & 0.270 & & & & & 0.677 & & LDA & \cite{pang2012mechanical} \\
			& & 0.486 & 0.171 & 0.393 & & & & & & & LDA & \cite{janisch2010ab} \\
			& & & 0.195 & 0.413 & & & & & & & LDA & \cite{wright1994density} \\
			& & & 0.190 & & & & & & & & LDA & \cite{thomson1997ab} \\
			& & & 0.206 & & & & & & & & LDA & \cite{thomson2000insight} \\
			\midrule
			Cu & 0.806 & 0.740 & 0.347 & 0.684 & 0.954 & 0.818 & 0.061 & 0.784 & 0.787 & 0.872 & PBE & Present \\
			& & & & 0.634 & & & 0.071 & & 0.856 & & PBE & \cite{zheng2020grain} \\
			& & 0.740 & 0.320 & 0.640 & & & 0.030 & & 0.800 & & PBE & \cite{hallberg2016investigation} \\
			& & & 0.320 & 0.640 & & & 0.020 & & 0.810 & & PBE & \cite{xu2016self} \\
			& & & & 0.570 & & & 0.020 & & & & PBE & \cite{bean2016origin} \\
			& & & & 0.558 & & & 0.020 & & & & PBE & \cite{nishiyama2020application} \\
			& & & & & & & 0.022 & & & & PBE & \cite{gao2014first} \\
			& & & 0.284 & 0.547 & & & 0.012 & 0.787 & & & PW91 & \cite{tsuru2010incipient} \\
			& & & & & & & 0.027 & & 0.755 & & QMAS & \cite{wang2009first} \\
			\midrule
			Ni & 1.131 & 1.016 & 0.443 & 0.943 & 1.193 & 1.185 & 0.074 & 1.157 & 1.211 & 1.138 & PBE & Present \\
			& & & & 1.150 & & & -0.007 & & 1.161 & & PBE & \cite{zheng2020grain} \\ 
			& & & 0.470 & & & & 0.090 & & & & PBE & \cite{chen2017role} \\
			& & & & 0.840 & & & 0.040 & & & & PBE & \cite{bean2016origin} \\
			& & & & & & & 0.045 & & & & PBE & \cite{gao2014first} \\
			& & & & & & & & & 1.236 & & PBE & \cite{pan2018development} \\
			& & & & 0.866 & & & & & & & PW91 & \cite{siegel2005computational} \\
			\midrule
			Ag & 0.484 & 0.425 & 0.171 & 0.382 & 0.561 & 0.482 & 0.009 & 0.452 & 0.402 & 0.481 & PBE & Present \\
			& & & & 0.427 & & & 0.069 & & 0.507 & & PBE & \cite{zheng2020grain} \\
			& & & & 0.339 & & & 0.012 & & & & PBE & \cite{nishiyama2020application} \\
			& & & & & & & 0.014 & & 0.441 & & PBE & \cite{pan2018development} \\
			\midrule
			Pd & 0.782 & 0.693 & 0.317 & 0.650 & 0.916 & 0.822 & 0.061 & 0.727 & 0.819 & 0.815 & PBE & Present \\
			& & & & 0.682 & & & 0.072 & & 0.852 & & PBE & \cite{zheng2020grain} \\
			& & 0.760 & 0.270 & 0.670 & & & 0.050 & 0.530 & 0.810 & & PBE & \cite{lovvik2018grain} \\
			& & & & 0.629 & & & 0.063 & & & & PBE & \cite{nishiyama2020application} \\		
			\midrule
			Au & 0.441 & 0.411 & 0.168 & 0.362 & 0.465 & 0.333 & 0.032 & 0.368 & 0.330 & 0.429 & PBE & Present \\
			& & & & 0.338 & & & 0.026 & & 0.393 & & PBE & \cite{zheng2020grain} \\
			& & & & 0.201 & & & 0.009 & & & & PBE & \cite{nishiyama2020application} \\
			& & & & & & & 0.010 & & & & PBE & \cite{o2018grain} \\
			\midrule
			Pt & 0.938 & 0.809 & 0.343 & 0.741 & 0.853 & 0.723 & 0.176 & 0.828 & 0.728 & 0.889 & PBE & Present \\
			& & & & 0.675 & & & 0.176 & & 0.795 & & PBE & \cite{zheng2020grain} \\
			& & & & 0.531 & & & 0.136 & & & & PBE & \cite{nishiyama2020application} \\
			& & & & & & & 0.118 & & & & PBE & \cite{o2018grain} \\
			\midrule
			Rh & 1.545 & 1.290 & 0.450 & 1.166 & 1.824 & 1.363 & 0.062 & 1.502 & 1.338 & 1.460 & PBE & Present \\
			& & & & 1.250 & & & 0.082 & & 1.311 & & PBE & \cite{zheng2020grain} \\
			\midrule
			Ir & 1.715 & 1.605 & 0.628 & 1.432 & 2.138 & 1.677 & 0.163 & 1.902 & 1.466 & 1.755 & PBE & Present \\
			& & & & 1.616 & & & 0.352 & & 1.578 & & PBE & \cite{zheng2020grain} \\
			\midrule
			Co & 1.088 & 0.988 & 0.421 & 0.921 & 1.533 & 1.301 & 0.041 & 1.034 & 1.287 & 1.182 & PBE & Present \\
			\bottomrule 
		\end{tabular*}\\}
\end{table}

\newpage
\section{Material properties}
\subsection{Lattice parameter, bulk modulus, elastic constants, and shear moduli}

Table~\ref{tableS3} shows the calculated lattice parameters, bulk modulus, elastic constants, and Voight average shear modulus for the studied fcc metals in the present work. Available theoretical (DFT~\cite{shang2010first} and EAM~\cite{holm2010comparing}) and experimental~\cite{simons1977single} data are also included for comparison. In general, the present results show a good agreement with other theoretical and experimental data.

\begin{table}[H]
	\renewcommand{\thetable}{S\arabic{table}}
	\caption{\label{tableS3}Comparison of lattice parameters ($a_{0}$), bulk modulus ($B_{0}$), elastic constants ($c_{11}$, $c_{12}$, $c_{44}$), and Voight average shear modulus ($\mu$) for fcc pure metals between the present and the previous works. All theoretical and experimental data correspond to the static conditions (0 K), except those experimental results at the room temperature marked by $\ast$.} 
	\begin{tabular*}{1.06\textwidth}{lllllllllllll}
		\toprule
		& Al & Au & Ag & Cu & Ni & Pd & Pt & Rh & Ir & Co & Functional & Ref. \\
		\midrule
		$a_{0}$ (\AA) & 4.04 & 4.16 & 4.15 & 3.64 & 3.52 & 3.94 & 3.97 & 3.82 & 3.87 & 3.52 & PBE-GGA & Present \\
		& 4.05 & 4.17 & 4.16 & 3.64 & 3.52 & 3.96 & 3.99 & 3.84 & 3.88 & 3.52 & PBE-GGA &~\cite{shang2010first} \\
		& 4.03 & 4.08 & & 3.62 & 3.52 & & & & & & EAM &~\cite{holm2010comparing} \\
		& 4.05 & 4.08 & 4.09 & 3.62 & 3.53 & 3.89 & 3.92 & 3.80 & 3.84 & & Expt. &~\cite{simons1977single} \\
		$B_{0}$ (GPa) & 78.2 & 138.0 & 90.1 & 140.4 & 195.6 & 167.4 & 248.2 & 257.5 & 350.0 & 208.4 & PBE-GGA & Present \\
		& 74.3 & 137.6 & 91.3 & 137.5 & 195.6 & 163.7 & 243.4 & 253.4 & 342.8 & 210.2 & PBE-GGA &~\cite{shang2010first} \\
		& 80.9 & 167.0 & & 138.3 & 180.4 & & & & & & EAM &~\cite{holm2010comparing} \\
		& 79.4 & 180.3 & 108.7 & 142.0 & 187.6 & 195.4 & 288.4 & 267.0 & 354.7 & & Expt. &~\cite{simons1977single} \\
		$c_{11}$ (GPa) & 108.9 & 154.7 & 108.2 & 177.0 & 274.2 & 199.3 & 309.5 & 411.6 & 586.2 & 291.4 & PBE-GGA & Present \\
		& 101.0 & 159.1 & 115.9 & 174.8 & 275.5 & 198.0 & 296.4 & 405.3 & 580.8 & 296.6 & PBE-GGA &~\cite{shang2010first} \\
		& 118.1 & 185.8 & & 169.9 & 240.5 & & & & & & EAM &~\cite{holm2010comparing} \\
		& 114.3 & 201.6 & 131.4 & 176.2 & 261.2 & 234.1 & 358.0 & 413.0$^{\ast}$ & 580.0$^{\ast}$ & & Expt. &~\cite{simons1977single} \\
		$c_{12}$ (GPa) & 62.8 & 129.7 & 81.1 & 122.1 & 156.3 & 151.4 & 217.5 & 180.5 & 231.9 & 166.9 & PBE-GGA & Present \\
		& 61.0 & 136.7 & 85.1 & 122.8 & 160.1 & 155.7 & 225.6 & 185.5 & 232.0 & 171.9 & PBE-GGA &~\cite{shang2010first} \\
		& 62.3 & 157.1 & & 122.6 & 150.3 & & & & & & EAM &~\cite{holm2010comparing} \\
		& 61.9 & 169.7 & 97.3 & 124.9 & 150.8 & 176.1 & 253.6 & 194.0$^{\ast}$ & 242.0$^{\ast}$ & & Expt. &~\cite{simons1977single} \\
		$c_{44}$ (GPa) & 32.6 & 23.3 & 43.0 & 75.5 & 131.2 & 57.7 & 58.4 & 185.1 & 258.9 & 142.6 & PBE-GGA & Present \\
		& 25.4 & 27.6 & 42.1 & 76.3 & 126.3 & 69.7 & 50.7 & 176.5 & 249.8 & 144.0 & PBE-GGA &~\cite{shang2010first} \\
		& 36.7 & 38.9 & & 76.2 & 119.2 & & & & & & EAM &~\cite{holm2010comparing} \\
		& 31.6 & 45.4 & 51.1 & 81.8 & 131.7 & 71.2 & 77.4 & 184.0$^{\ast}$ & 256.0$^{\ast}$ & & Expt. &~\cite{simons1977single} \\
		$\mu$ (GPa) & 28.8 & 19.0 & 31.2 & 56.3 & 102.3 & 44.2 & 53.3 & 157.3 & 226.2 & 110.5 & PBE-GGA & Present \\
		& 23.2 & 21.1 & 31.4 & 56.2 & 98.9 & 50.3 & 44.6 & 149.9 & 219.6 & 111.3 & PBE-GGA &~\cite{shang2010first} \\
		& 33.0 & 29.1 & & 55.2 & 90.0 & & & & & & EAM &~\cite{holm2010comparing} \\
		& 29.4 & 33.6 & 37.5 & 59.3 & 101.1 & 54.3 & 67.3 & 154.2$^{\ast}$ & 221.2$^{\ast}$ & & Expt. &~\cite{simons1977single} \\
		\bottomrule
	\end{tabular*}\\
\end{table}

\subsection{Surface energy}
For surface energy, extensively theoretical works have been employed for fcc metals~\cite{lee2018surface, patra2017properties, lin2020compensation, tran2016surface, swart2007surface, yoo2016exploring, wang2014surface, wu2010ab}. In Table~\ref{tableS4}, we collect several works using different exchange-correlation functionals. For using the same exchange-correlation functional, it can be seen that the calculated surface energies show a good agreement with each other. To be consistent, we adopt the surface energies from the work with results for most fcc metals~\cite{lee2018surface} and the other one for Co is taken from Ref.~\cite{swart2007surface}. The corresponding values are in bold. 

\begin{table}[H]
	\renewcommand{\thetable}{S\arabic{table}}
	\caption{\label{tableS4}Surface energies of fcc metals. Those adopted in the present work are in bold.}
	\begin{tabular*}{0.93\textwidth}{lllllllllllll}
		\toprule
		Surface & Al & Au & Ag & Cu & Ni & Pd & Pt & Rh & Ir & Co & Functional & Ref. \\
		\midrule
		(100) & \textbf{0.92} & \textbf{0.86} & \textbf{0.84} & \textbf{1.44} & \textbf{2.22} & \textbf{1.51} & \textbf{1.85} & \textbf{2.35} & \textbf{2.84} & & PBE-GGA &~\cite{lee2018surface} \\
		& 0.95 & 0.86 & 0.81 & 1.48 & & 1.79 & 1.88 & 2.77 & & & PBE-GGA &~\cite{patra2017properties} \\
		& & & & & 2.18 & 1.51 & & 2.32 & 2.80 & 2.45 & PBE-GGA &~\cite{lin2020compensation} \\
		& & & & & 2.21 & & 1.84 & & & & PBE-GGA &~\cite{tran2016surface} \\
		& & 0.90 & & 1.46 & & & & 2.34 & & \textbf{2.46} & PBE-GGA &~\cite{swart2007surface} \\
		& 1.15 & 1.39 & 1.16 & 1.99 & & 2.43 & 2.35 & 3.04 & & & LDA-GGA &~\cite{patra2017properties} \\
		& & 1.33 & & 2.02 & & & & 2.95 & & & LDA-GGA &~\cite{swart2007surface} \\
		& 1.08 & 1.13 & 1.04 & 1.76 & & 2.15 & 2.21 & 2.97 & & & PBEsol-GGA &~\cite{patra2017properties} \\
		& 1.04 & 1.19 & 1.14 & 1.81 & 2.61 & 1.84 & 2.23 & 2.84 & 3.24 & & PBEsol-GGA &~\cite{yoo2016exploring}\\
		& & 1.36 & 1.27 & 2.15 &  & 2.15 & 2.47 & 3.49 & & & PBE-GGA &~\cite{wang2014surface} \\
		\midrule
		(111) & \textbf{0.82} & \textbf{0.71} & \textbf{0.76} & \textbf{1.30} & \textbf{1.92} & \textbf{1.33} & \textbf{1.49} & \textbf{2.01} & \textbf{2.06} & & PBE-GGA &~\cite{lee2018surface} \\
		& 0.77 & 0.75 & 0.78 & 1.33 & & 1.36 & 1.56 & 2.09 & & & PBE-GGA &~\cite{patra2017properties} \\
		& & & & & 1.91 & 1.33 & & 2.00 & 2.28 & 2.07 & PBE-GGA &~\cite{lin2020compensation} \\
		& 0.71 & 0.71 & 0.74 & 1.29 & 1.93 & 1.36 & 1.51 & 2.02 & 2.31 & & PBE-GGA &~\cite{wu2010ab} \\
		& & 0.76 & & & 1.92 & 1.33 & 1.48 & & 2.42 & & PBE-GGA &~\cite{tran2016surface} \\
		& & 0.70 & & 1.32 & & & & 2.02 & & \textbf{2.02} & PBE-GGA &~\cite{swart2007surface} \\
		& 0.99 & 1.24 & 1.13 & 1.81 & & 1.88 & 1.98 & 2.67 & & & LDA-GGA &~\cite{patra2017properties} \\
		& & 1.11 & & 1.85 & & & & 2.46 & & & LDA-GGA &~\cite{swart2007surface} \\
		& 0.99 & 1.10 & 1.00 & 1.59 & & 1.63 & 1.85 & 2.40 & & & PBEsol-GGA &~\cite{patra2017properties} \\
		& 0.93 & 1.01 & 1.07 & 1.67 & 2.28 & 1.59 & 1.81 & 2.44 & 2.56 & & PBEsol-GGA &~\cite{yoo2016exploring}\\
		& & 1.14 & 1.15 & 1.94 &  & 1.90 & 2.00 & 2.78 & & & PBE-GGA &~\cite{wang2014surface} \\
		\bottomrule
	\end{tabular*}\\
\end{table}

\newpage
\section{The correlation between GBEs in different metals}

\begin{figure}[H]\setcounter{subfigure}{0}
	\centering
	\renewcommand{\thefigure}{S\arabic{figure}}
	\subfigure[]{\includegraphics[scale=0.5]{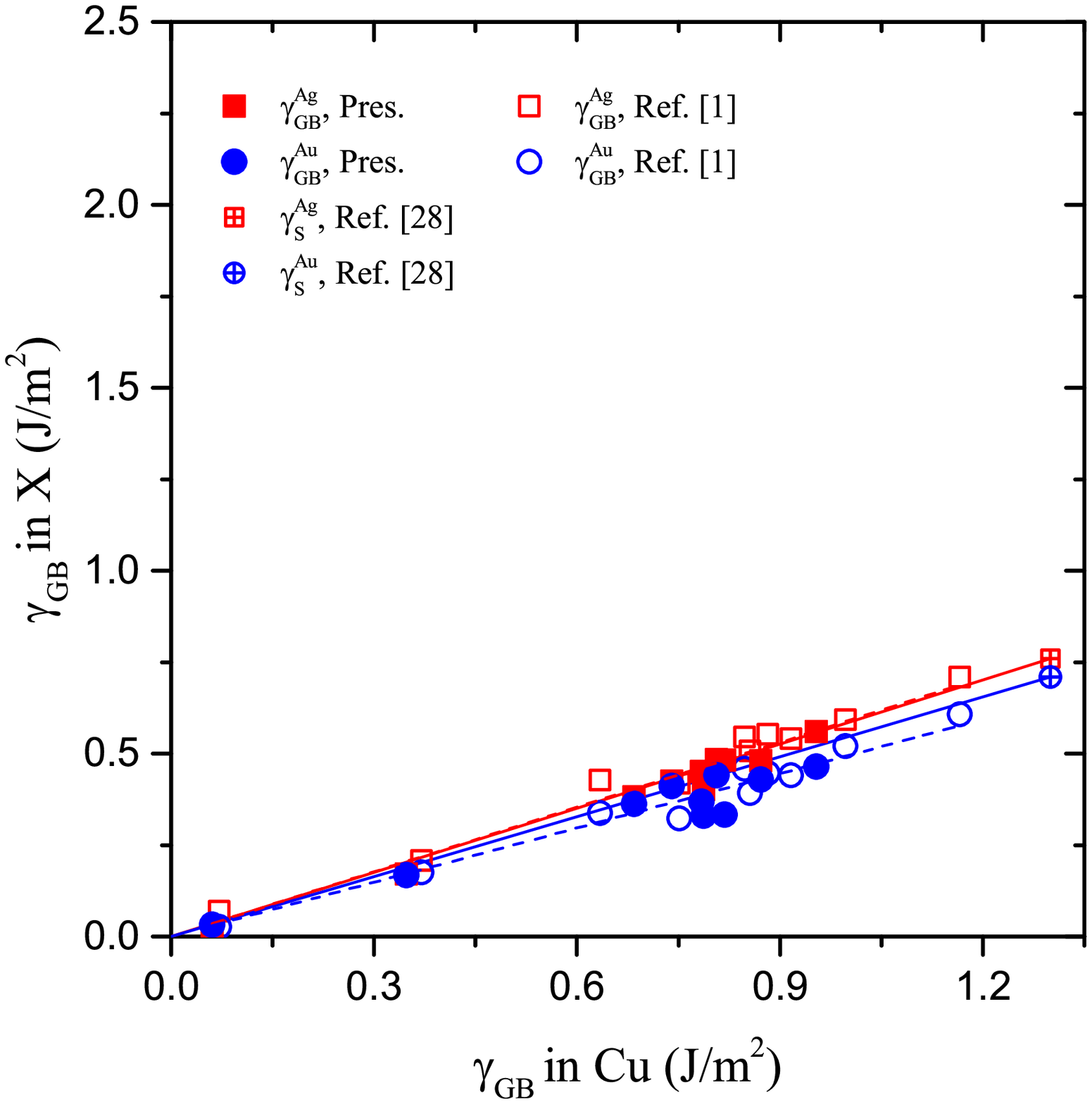}}
	\subfigure[]{\includegraphics[scale=0.5]{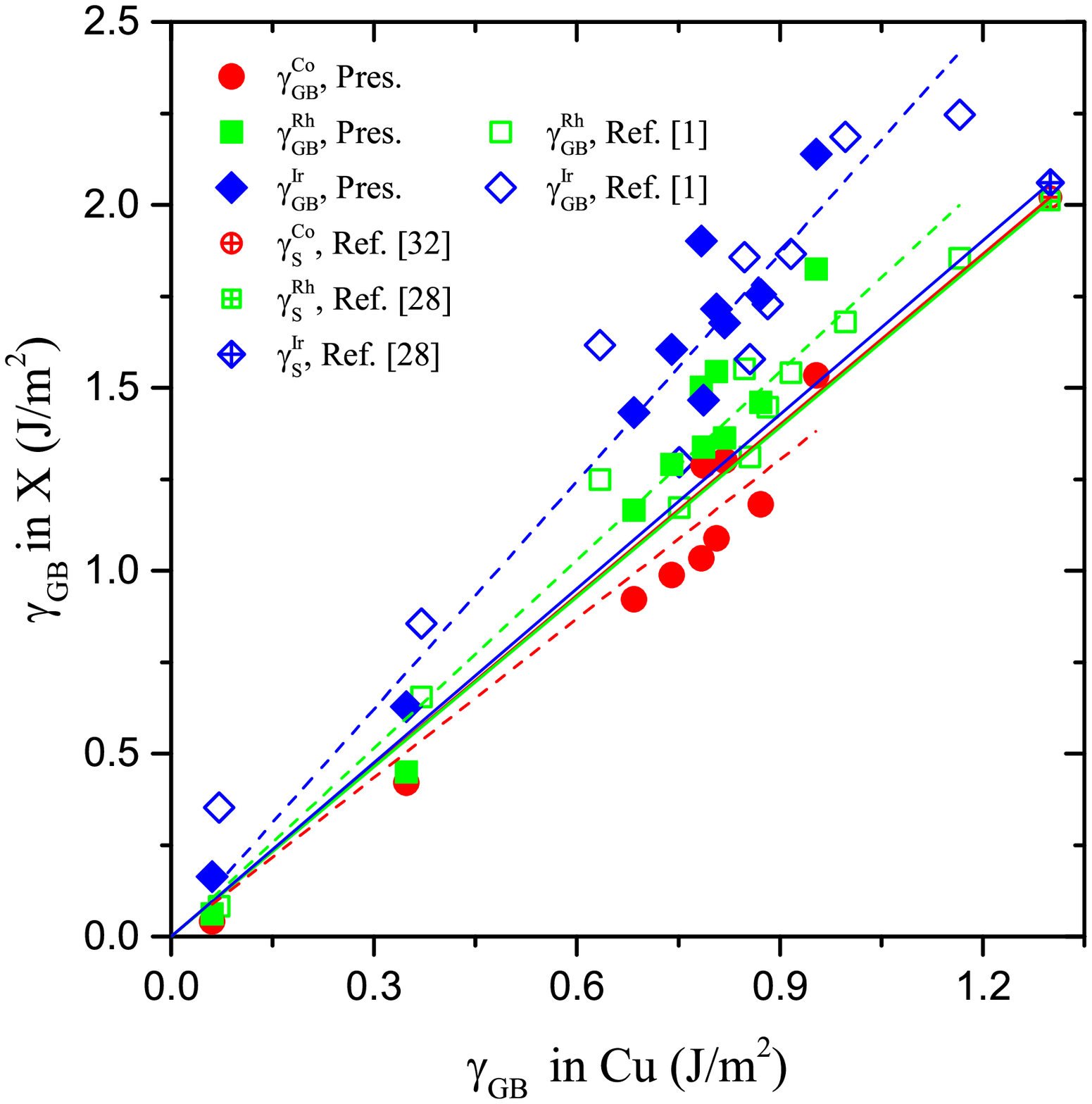}}
	\caption{\label{figS2}(Color online) Pairwise comparison of the calculated $\gamma_{\rm GB}$ (solid symbols) and the (111) surface energies (open symbols with plus) for Ag, Au, Co, Rh, and Ir with Cu obtained at 0 K. For clearance, the results are shown in two subfigures. The dashed lines are the linear fit of the GBEs. Previous DFT GBEs (open symbols) for tilt and twist GBs from Ref.~\cite{zheng2020grain} are included in the linear fitting. DFT surface energies are taken from Refs.~\cite{lee2018surface, swart2007surface}.}
\end{figure}

Fig.~\ref{figS2} shows the pairwise comparison of the calculated GBEs and the (111) surface energies for the remaining elements, Ag, Au, Co, Rh, and Ir with Cu. Previous DFT GBEs (solid symbols) for tilt and twist GBs from Ref.~\cite{zheng2020grain} are included in the linear fitting. DFT surface energies are taken from Refs.~\cite{lee2018surface, swart2007surface}.

\begin{table}[H]
	{\small
		\renewcommand{\thetable}{S\arabic{table}}
		\caption{\label{tableS5}The pairwise comparison of the ratios of $a_{0}c_{44}$, $a_{0}\mu$, surface energies ($\gamma_{\rm S}$), and $\delta_{\rm GB(fit)}^{\rm A/Cu}$ in fcc metals. The EAM, experimental, and VASP results are from Refs.~\cite{holm2010comparing},~\cite{simons1977single}, and~\cite{zheng2020grain, lee2018surface, swart2007surface}, respectively.}
		\begin{tabular*}{1.02\textwidth}{l|lll|lll|ll|ll|ll}
			\toprule
			\multirow{2}*{A} & \multicolumn{3}{c}{$\delta_{a_{0}c_{44}}^{\rm A/Cu}$} & \multicolumn{3}{c}{$\delta_{a_{0}\mu}^{\rm A/Cu}$} & $\delta_{\rm S(100)}^{\rm A/Cu}$ & $\delta_{\rm S(111)}^{\rm A/Cu}$ & $\delta_{\rm GB(fit)}^{\rm A/Cu}$ & $R^{2}$ & $\delta_{\rm GB(fit)}^{\rm A/Cu}$ & $R^{2}$ \\
			& (Pres.) & (EAM) & (Expt.) & (Pres.) & (EAM) & (Expt.) & (VASP) & (VASP) & (VASP) & (VASP) & (Pres.) & (Pres.) \\
			\midrule
			Au & 0.353 & 0.576 & 0.626 & 0.386 & 0.595 & 0.639 & 0.597 & 0.546 & 0.501 & 0.995 & 0.486 & 0.989 \\ 
			Ag & 0.650 & & 0.706 & 0.633 & & 0.714 & 0.583 & 0.585 & 0.607 & 0.997 & 0.568 & 0.997 \\ 
			Al & 0.480 & 0.537 & 0.432 & 0.569 & 0.667 & 0.555 & 0.639 & 0.631 & 0.540 & 0.991 & 0.559 & 0.980 \\ 
			Pd & 0.829 & & 0.935 & 0.852 & & 0.984 & 1.049 & 1.023 & 1.014 & 0.999 & 0.964 & 0.998 \\ 
			Pt & 0.845 & & 1.025 & 1.037 & & 1.228 & 1.285 & 1.146 & 0.990 & 0.991 & 1.006 & 0.987 \\ 
			Co & 1.828 & &  & 1.899 & &  & 1.708 & 1.554 &  &  & 1.448 & 0.990 \\ 
			Ni & 1.685 & 1.523 & 1.570 & 1.762 & 1.588 & 1.661 & 1.542 & 1.477 & 1.441 & 0.994 & 1.384 & 0.995 \\ 
			Rh & 2.581 & & 2.361 & 2.941 & & 2.728 & 1.632 & 1.546 & 1.667 & 0.995 & 1.714 & 0.994 \\ 
			Ir & 3.655 & & 3.320 & 4.282 & & 3.954 & 1.972 & 1.585 & 2.036 & 0.988 & 2.119 & 0.994 \\
			\bottomrule 
		\end{tabular*}\\}
\end{table}

\newpage
\section{Temperature dependence of the experimental GBEs}
\begin{figure}[h!]
	\centering
	\renewcommand{\thefigure}{S\arabic{figure}}
	\includegraphics[scale=0.6]{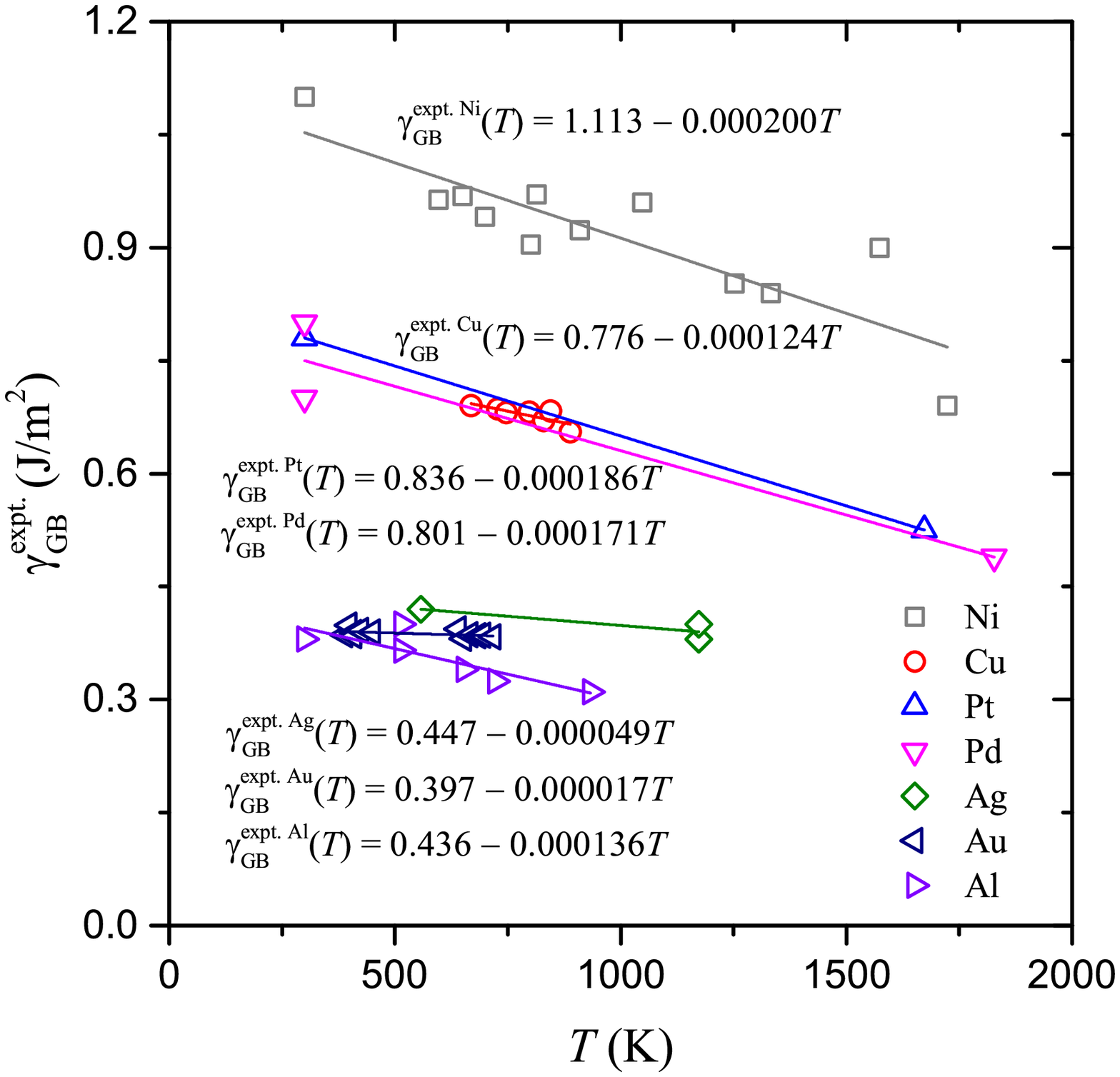}
	\caption{\label{figS3}(Color online) Temperature dependence of the mean (experimental) GBEs ($\gamma_{\rm GB}^{\rm expt.}$) in fcc metals. The symbols represent the measured $\gamma_{\rm GB}^{\rm expt.}$ of the high-angle grain boundaries in Al~\cite{murr1973twin, bolling1968average, prokofjev2019estimation}, Ni~\cite{divinski2010grain, murr1970interfacial, miedema1978surface}, Cu~\cite{gupta1977influence}, Pd~\cite{miedema1978surface, birringer2002interface, birringer2003interface, prokofjev2019estimation}, Ag~\cite{bolling1968average, prokofjev2019estimation}, Pt~\cite{mclean1971study, hondors1982metallic}, and Au~\cite{gupta1977influence}. The solid lines represent the linear fit results.}
\end{figure}

Experimentally, the mean GBE for high angle GBs can be obtained at elevated temperatures. As shown in Fig.~\ref{figS3}, the mean (experimental) GBEs for different metals are collected~\cite{murr1973twin, bolling1968average, prokofjev2019estimation, divinski2010grain, murr1970interfacial, miedema1978surface, gupta1977influence, birringer2002interface, birringer2003interface, mclean1971study, hondors1982metallic}. Gupta~\textit{et al.}~\cite{gupta1977influence} measured the GBEs of Au and Cu at different temperatures and obtained the corresponding temperature coefficients through the linear fitting. To extrapolate the experimental GBEs to 0 K or room temperature, linear fitting based on the measured GBEs at elevated temperatures is used. The corresponding linear function and temperature coefficient for pure metals are shown in Fig.~\ref{figS3}.

\newpage
\section{Estimation of $\delta_{\rm S}^{\rm A/Cu}$ at room temperature.}

Following the work by Tyson~\textit{et al.}~\cite{tyson1977surface}, we can estimate the change of $\delta_{\rm S}^{\rm A/Cu}$ as we go from 0 K to room temperature. 
Tyson~\textit{et al.} computed the surface energy at low temperature using the surface tension data in the liquid phase and the linear temperature dependence $\gamma_{\rm S}(0)-\gamma_{\rm S}(T_{m})\approx RT_{m}/A$, connecting the surface energy at melting temperature $T_{m}$ and at 0 K. Here $R$ is the gas constant and $A$ is the surface area per mole of surface atoms. Writing this expression for temperature $T$ and taking the difference, we arrive at $\gamma_{\rm S}(T)-\gamma_{\rm S}(0)\approx -\alpha \cdot T/T_{m}$, where $\alpha=RT_{m}/A$. 
Thus the temperature dependence of $\delta_{\rm S}^{\rm A/Cu}(T)$ in leading order in $T$ can be written as 
\begin{equation}
	\delta_{\rm S}^{\rm A/Cu}(T)\approx \delta_{\rm S}^{\rm A/Cu}(0)+\left [\frac{1}{\gamma_{\rm S}^{\rm Cu}(0)} \left (\frac{\alpha^{\rm Cu}}{T_{m}^{\rm Cu}}\cdot \frac{\gamma_{\rm S}^{\rm A}(0)}{\gamma_{\rm S}^{\rm Cu}(0)}-\frac{\alpha^{\rm A}}{T_{m}^{\rm A}} \right )\right ]\cdot T,
\end{equation}
where the $\gamma_{\rm S}^{\rm Cu}(0)$ and $\gamma_{\rm S}^{\rm A}(0)$ are the 0 K surface energies for Cu and metal A, respectively. 
Using the $\alpha$ and $\gamma_{\rm S}(0)$ values for pure metals by Tyson~\textit{et al.}~\cite{tyson1977surface}, we obtain the change of $\delta_{\rm S}^{\rm A/Cu}$ when going from 0 K to room temperature. 
The results for the present metals are listed in Table~\ref{tableS6}. The average change of $\Delta \delta_{\rm S}^{\rm A/Cu}$ is $\sim$0.7\%, which implies that the ratio of surface energies changes very weakly when increasing the temperature to 298 K.

\begin{table}[H]
	\centering
	\renewcommand{\thetable}{S\arabic{table}}
	\caption{\label{tableS6}Calculated $\delta_{\rm S}^{\rm A/Cu}$ values for the selected fcc metals at 0 K and 298 K.}
	\begin{tabular*}{0.4\textwidth}{lllll}
		\toprule
		A & $\delta_{\rm S}^{\rm A/Cu}(0)$ & $\delta_{\rm S}^{\rm A/Cu}(298)$ & $\Delta \delta_{\rm S}^{\rm A/Cu}$ & \% \\
		\midrule
		Al & 0.640 & 0.636 & 0.004 & 0.6 \\
		Au & 0.843 & 0.842 & 0.001 & 0.1 \\
		Ag & 0.697 & 0.695 & 0.002 & 0.3 \\
		Ni & 1.331 & 1.324 & 0.007 & 0.5 \\
		Pd & 1.125 & 1.118 & 0.007 & 0.6 \\
		Pt & 1.398 & 1.383 & 0.015 & 1.1 \\
		Co & 1.410 & 1.400 & 0.010 & 0.7 \\
		Rh & 1.494 & 1.478 & 0.016 & 1.1 \\
		Ir & 1.712 & 1.689 & 0.023 & 1.3 \\
		\bottomrule
	\end{tabular*}\\
\end{table}






\newpage
\bibliographystyle{apsrev4-2}
\bibliography{References.bib}